\documentclass[conference]{IEEEtran}
\IEEEoverridecommandlockouts
\usepackage{cite}
\usepackage{amsmath,amssymb,amsfonts}
\usepackage{algorithmic}
\usepackage{graphicx}
\usepackage{textcomp}
\usepackage{xcolor}
\def\BibTeX{{\rm B\kern-.05em{\sc i\kern-.025em b}\kern-.08em
    T\kern-.1667em\lower.7ex\hbox{E}\kern-.125emX}}
\begin{document}

\title{Variable-Rate Deep Image Compression based on Low-Rank Adaptation by Progressive Learning}

\author{
\IEEEauthorblockN{Xing-Yu Xu}
\IEEEauthorblockA{\textit{Communications and Multimedia Lab} \\
\textit{National Taiwan University}\\
Taipei, Taiwan \\
nick19960723@cmlab.csie.ntu.edu.tw}
\and
\IEEEauthorblockN{Chen-Hsiu Huang}
\IEEEauthorblockA{\textit{Communications and Multimedia Lab} \\
\textit{National Taiwan University}\\
Taipei, Taiwan \\
chenhsiu48@cmlab.csie.ntu.edu.tw}
\and
\IEEEauthorblockN{Ja-Ling Wu}
\IEEEauthorblockA{\textit{Communications and Multimedia Lab} \\
\textit{National Taiwan University}\\
Taipei, Taiwan \\
wjl@cmlab.csie.ntu.edu.tw}
}

\maketitle

\begin{abstract}
In the digital age, image compression is crucial for numerous applications, including web media, streaming services, high-resolution medical imaging, and connected vehicle networks, enabling efficient data storage and transmission. With the increasing demand for high-quality image communication, the need for advanced compression techniques becomes increasingly critical. Numerous  Deep Image Compression (DIC) techniques have recently been introduced, showing impressive performance compared to traditional standards. However, variable-rate image compression remains an unresolved issue. Specific DIC methods deploy multiple networks to attain different compression rates, whereas others use a single model, which often results in higher computational complexity and reduced performance. This work proposes a progressive learning approach for variable-rate image compression based on the parameter-efficient fine-tuning method, the Low-Rank Adaptation (LoRA). We introduce an additional LoRA Rate-Adaptive Module (LoRAM) in DIC methods. Due to the re-parameterized merging of LoRA, our proposed method does not introduce additional computational complexity during inference. Compared to methods utilizing multiple models, comprehensive experiments demonstrate that our approach achieves competitive performance, saving 99\% in parameter storage, 90\% in datasets, and 97\% in training steps.
\end{abstract}


\section{Introduction}
For decades, image compression has been a focal point of research and has recently surged in significance with the widespread adoption of mobile devices for image capturing and sharing. In the course of recent decades, numerous image compression standards have been put forth and extensively developed, including JPEG \cite{b23}, BPG \cite{b4}, and VVC \cite{b5}. Nonetheless, the emergence of deep learning has ushered in a new era termed deep image compression, which offers notable advancements over traditional methods. Deep learning approaches harness neural networks to acquire data-centric, refined compression algorithms, frequently attaining superior compression rates while maintaining minimal perceptible degradation in image quality.

A fundamental aspect of deep image compression lies in the rate-distortion function, which is grounded in information theory \cite{b25}. This function effectively manages the trade-off between compression rate (the amount of retained data) and distortion (the loss of quality due to compression). A solitary hyper-parameterized Lagrange multiplier \(\lambda\) corresponds to a conventional configuration compression model with a fixed bitrate. Some Variational Autoencoder (VAE) based image compression techniques necessitate training multiple fixed-rate models to achieve rate adaptation, with each model dedicated to a specific rate. Consequently, the training cost and memory requirements escalate significantly as the desired rate range expands and refines. 

The conditional autoencoder \cite{b7, b27}, employs fully connected layers into the convolutional unit, enabling discrete rate adaptation but incurring extra computational complexity and memory requirements. The introduction of mixed bin sizes \cite{b7} aimed to expand coverage from discrete points to a broader continuous rate range, yet resulted in degraded rate-distortion performance. The bottleneck scaling scheme \cite{b28,b1,b8} disregards compatibility between the autoencoder, scaling parameters, and image spatial dependence, resulting in poor low bit rate range performance.

LoRA \cite{b12} has demonstrated significant potential as an efficient fine-tuning paradigm. The fine-tuning process involves adding an exceedingly small number of additional parameters to yield comparable performance to that of fully trained models. To maintain comparable rate-distortion performance while reducing additional model parameters and computational complexity, we propose to add the LoRA Rate-Adaptive Module (LoRAM) in existing deep compression methods. LoRAM utilizes extremely few additional model parameters during training to fine-tune the backbone network. It merges with the backbone network during inference (compression/decompression) without introducing additional latency. The parameters of LoRAM constitute merely 1\% of those in the backbone network, yet it yields performance on par with independently trained models. Additionally, our method attains favorable outcomes with just 10\% of the data during LoRA-tuning, showcasing remarkable sample efficiency. Furthermore, unlike models trained individually for various compression rates, our approach converges more swiftly, requiring less time and saving 97\% in training steps.

\section{Related Work}

\subsection{Deep Image Compression}

Deep image compression (DIC) has become a powerful method, utilizing Deep Neural Networks (DNN) to minimize distortion between original and reconstructed images while optimizing the probability of quantized representations at low entropy coding costs (bit rate). Deep image compression generally entails a balance between compression rate and distortion, managed through hyper-parameterized Lagrange multipliers \( \lambda \) in the rate-distortion loss function. Each Lagrange multiplier corresponds to a particular model crafted to balance different trade-offs between compression rates and image quality. Recent advancements in learned image compression primarily employ deep neural networks, exploring a variety of model architectures such as recurrent networks \cite{b30,b22} and autoencoders featuring an entropy-constrained bottleneck \cite{b2,b3,b15,b16,b19,b20}. Incorporating attention mechanisms simulates the cognitive processes observed in biological systems, allocating attentional resources more effectively to relevant regions for capturing intricate details and suppressing irrelevant information. The non-local attention mechanism \cite{b31} has demonstrated advantages across various visual tasks. In the realm of learned image compression, \cite{b17} employs a non-local attention module to create an implicit importance matrix, which directs latent features' spatial and channel-wise adaptive processing. In contrast, \cite{b6} streamlines the attention mechanism by excluding the non-local block. Furthermore, \cite{b32} employs a window block with window-based attention to replace the non-local block, maintaining comparable rate-distortion performance while reducing computational complexity. In addition, Symmetrical TransFormer (STF)\cite{b32} utilizes the Swin-Transformer \cite{b18} alongside sub-pixel convolution to effectively capture long-range spatial redundancy.

\subsection{Variable-Rate Deep Image Compression}

Choi et al. \cite{b7} developed a VAE-based model with an additional autoencoder. However, uncertainty surrounding discrete rank and bin size compromises model efficacy, presenting challenges in selecting optimal combinations within neighboring coverage areas. However, determining the discrete levels and bin sizes is not straightforward, leading to suboptimal quality in the output images. \cite{b27} achieves continuous modulation of compression rates by interpolating learned parameters within a pre-trained discrete variable-rate model. These approaches typically employ channel-wise affine transforms conditioned on Lagrange multipliers, demonstrating accuracy comparable to independently trained single-rate models. Additionally, incorporating a fully connected layer contributes to a heavier network. Asymmetric Gained Variational Autoencoder (AG-VAE), as introduced in \cite{b8}, employs asymmetric gain units to enable discrete rate adaptation within a unified model. This approach exhibits performance on par with models trained independently by utilizing channel affine transformations. The work presented in \cite{b26} introduces a flexible deep image compression network rooted in spatial feature transformation. The network produces compressed images at variable rates using  a source image and its corresponding quality map as input.  Qin et al. \cite{b24} present a progressive learning method that introduces prompt tuning for variable-rate image compression, which involves training separately for different target compression rates.

\subsection{Low-Rank Adaptation}

In the LoRA technique introduced by \cite{b12}, two trainable low-rank matrices are employed to update network parameters. Within LoRA, a down-projection matrix and an up-projection matrix operate as paired components alongside the Q, K, and V matrices, representing the query, key, and value within the attention layer of the transformer. After obtaining a pre-trained weight matrix \( W \in \mathbb{R}^{d \times k} \). During training, LoRA updates weights through low-rank decomposition, as expressed by \( \Delta W = W_{\text{down}} \cdot W_{\text{up}} \). While the weights of the Pretrained Language Models (PLMs) remain frozen, only the low-rank matrices of LoRA, namely \( W_{\text{down}} \in \mathbb{R}^{d \times r} \) and \( W_{\text{up}} \in \mathbb{R}^{r \times k} \), are fine-tuned (\( r \ll \{d, k\} \)). In the inference stage, the LoRA weights seamlessly merge with the original weight matrix of the PLMs, maintaining the same inference time. This integration incorporates a scaling factor (\( s = 1/r \)) into the LoRA module.

\section{Proposed Method}

\subsection{Overview}

Since our approach is based on two architectures proposed by Zou et al. \cite{b32}, Window-Attention CNN (WACNN) and Symmetrical TransFormer (STF), we first overview the primary pipelines first. The WACNN architecture, as outlined in \cite{b32}, adopts the design from \cite{b21} for convenient comparison in subsequent analyses. W-MSA \cite{b18} and SW-MSA \cite{b18} are vital components of the STF \cite{b32}, representing window-based multi-head self-attention and shifted window-based self-attention, respectively. In addition to the baseline network, our method integrates the LoRA Rate-Adaptive Module (LoRAM) into each linear layer of the three components: encoder, decoder, and entropy model. The encoder \( \mathcal{G}_a \) takes an input image \( x \) and generates a corresponding latent representation \( y \). Upon quantization with \( Q \), let \( \hat{y} \) denote the discretized version of \( y \). Subsequently, the decoder \( \mathcal{G}_s \) reconstructs the image \( \hat{x} \) from the quantized representation \( \hat{y} \). The primary process can be formulated as follows:
\begin{equation}
    \begin{aligned}
        &y = \mathcal{G}_a(x;\phi), \\
        &\hat{y} = Q(y), \\
        &\hat{x} = \mathcal{G}_s(\hat{y};\theta),
    \end{aligned}    
\end{equation}
Here, \( \phi \) and \( \theta \) denote the trainable parameters of the decoder \( \mathcal{G}_s \) and encoder \( \mathcal{G}_a \), respectively. Quantization \( Q \) inevitably leads to clipping errors \( y - Q(y) \) in the latent representation, distorting the reconstructed image. In line with previous studies \cite{b21}, we mitigate quantization error by applying rounding and integrating the predicted quantization error during the training phase. We represent each element \( \hat{y} \) as a single Gaussian distribution characterized by its standard deviation \( \sigma_i \) and mean \( \mu_i \), with the introduction of additional side information \( \hat{z} \). The distribution \( p(\hat{y} | \hat{z}) \) of \( \hat{y} \) is modeled using an entropy model based on the Single Gaussian Model:
\begin{equation}
    \begin{aligned}
        &p_{\hat{y}|\hat{z}}(\hat{y}|\hat{z}) = \mathcal{N}(\mu_i, \sigma_i^2), \\
        &\mu, \sigma = \mathcal{G}_{ep}(\mathcal{H}_s(\hat{z};\psi); \xi), \\
        &\hat{z} = Q(\mathcal{H}_a(y;\omega)),
    \end{aligned}    
\end{equation}
where \( \omega \), \( \psi \) and \( \xi \) represent the trainable parameters of the hyper-encoder \( \mathcal{H}_s \), hyper-decoder \( \mathcal{H}_a \) and entropy parameter predictor \( \mathcal{G}_{ep} \) respectively. The loss function of the image compression model is:
\begin{equation}
    \begin{aligned}
        \mathcal{L} &= R + \lambda \cdot D \\
                    &= \mathbb{E}_{x\sim p_x}[-log_2 p_{\hat{y}|\hat{z}}(\hat{y}|\hat{z}) -log_2 p_{\hat{z}}(\hat{z})] \\ &\quad + \lambda \cdot \mathbb{E}_{x\sim p_x}[d(x, \hat{x})].
    \end{aligned}    
\end{equation}
Here, \( \lambda \) regulates the balance between rate and distortion, \( R \) signifies the bit rate of latent representations \( \hat{y} \) and hyper latent representations \( \hat{z} \), and \( d(x, \hat{x}) \) indicates the distortion between the original image \( x \) and the reconstructed image \( \hat{x} \).

\subsection{LoRA Rate-Adaptive Module}

Given a pre-trained weight matrix \( \bold{W}^k \in \mathbb{R}^{c_{out} \times c_{in}} \) for a specific layer, our objective is to adapt it for variable-rate deep image compression via efficient weight modification.
\begin{equation}
    \begin{aligned}
        \bold{h}^{k+1} &= \bold{W}^k\ \bold{h}^k + \Delta \bold{W}^k \bold{h}^k \\
                        &= (\bold{W}^k + \Delta \bold{W}^k)\ \bold{h}^k,
    \end{aligned}
\end{equation}
where \( \bold{h}^k \) and \( \bold{h}^{k+1} \) represent the \( k \)-th and \( (k + 1) \)-th layer of features. We utilize the low-rank property of \( \Delta \bold{W}^k \) to efficiently update instance-specific parameters, aiming to maximize reconstruction quality while minimizing bit overhead. We keep \( \bold{W}^k \) frozen and iteratively update it with two trainable matrices \( \bold{A} \) and \( \bold{B} \) through low-rank decomposition:
\begin{equation}
    \begin{aligned}
        \Delta \bold{W}^k = \bold{B}^k \bold{A}^k,
    \end{aligned}
\end{equation}
where \( \bold{A}^k \in \mathbb{R}^{r \times c_{in}} \) is randomly obtained from Gaussian distribution, and \( \bold{B}^k \in \mathbb{R}^{c_{out} \times r} \) is set to zeros during initialization. This initialization ensures \( \Delta \bold{W}^k = 0 \) at the beginning of adaptation. The rank \( r \ll \min\{c_{in}, c_{out}\} \). We can merge \( \Delta \bold{W}^k \) into \( \bold{W}^k \) to avoid introducing additional inference latency during inference.
We denote \( \Delta \phi = \{\bold{B}_{\phi} \), \( \bold{A}_{\phi}\} \), \( \Delta \omega = \{\bold{B}_{\omega} \), \( \bold{A}_{\omega}\} \), \( \Delta \theta = \{\bold{B}_{\theta} \), \( \bold{A}_{\theta}\} \), \( \Delta \psi = \{\bold{B}_{\psi} \), \( \bold{A}_{\psi}\}\) and \( \Delta \xi = \{\bold{B}_{\xi} \), \( \bold{A}_{\xi}\}\) presenting the LoRAM for the encoder, decoder, hyper-encoder and hyper-decoder, respectively. For various target compression rates, we train different \( \Delta \phi_i = \{\bold{B}_{\phi_i} \), \( \bold{A}_{\phi_i}\} \), \( \Delta \omega_i = \{\bold{B}_{\omega_i} \), \( \bold{A}_{\omega_i}\} \), \( \Delta \theta_i = \{\bold{B}_{\theta_i} \), \( \bold{A}_{\theta_i}\} \), \( \Delta \psi_i = \{\bold{B}_{\psi_i} \), \( \bold{A}_{\psi_i}\}\), and \( \Delta \xi_i = \{\bold{B}_{\xi_i} \), \( \bold{A}_{\xi_i}\}\) with different Lagrange multipliers \( \lambda_i \).

\subsection{Window Attention CNN with LoRA}
The Window Attention Module in WACNN is a crucial component for capturing long-range correlations and eliminating long-range redundancy. The feature map is divided into windows of size M × M without overlap. Attention maps are then computed separately within each window. \( X^{k}_{i} \) and \( X^{k}_{j} \) denote the i-th and j-th elements in the k-th window, respectively, as depicted below:
\begin{equation}\label{eq6}
    \begin{aligned}
        \bold{Y}^{k}_{i} = \frac{1}{C(\bold{X}^{k})}\sum_{\forall j} f(\bold{X}^{k}_{i}, \bold{X}^{k}_{j}) g(\bold{X}^{k}_{j}),
    \end{aligned}
\end{equation}
\begin{equation}
    \begin{aligned}
        \text{with}\ f(\bold{X}^{k}_{i}, \bold{X}^{k}_{j}) &= e^{\theta(\bold{X}^{k}_{i})^{T} \phi(\bold{X}^{k}_{j})}, \\
        C(\bold{X}^{k}) &= \sum_{\forall j} f(\bold{X}^{k}_{i}, \bold{X}^{k}_{j}).
    \end{aligned}
\end{equation}

In this formulation, \( \theta(\bold{X}^{k}_{i}) = \bold{W}_{\theta} \bold{X} + \Delta \bold{W}_{\theta} \bold{X} \), \( \phi(\bold{X}^{k}_{i}) = \bold{W}_{\phi} \bold{X} + \Delta \bold{W}_{\phi} \bold{X} \) and \( g(\bold{X}^{k}_{j}) = \bold{W}_{g}\bold{X}^{k}_{j} + \Delta \bold{W}_{g}\bold{X}^{k}_{j} \), where \( \bold{W}_{\theta} \), \( \bold{W}_{\phi} \), \( \bold{W}_{g} \), \( \Delta \bold{W}_{\theta} = \bold{B}_{\theta}\bold{A}_{\theta} \), \( \Delta \bold{W}_{\phi} = \bold{B}_{\phi}\bold{A}_{\phi} \), and \( \Delta \bold{W}_{g} = \bold{B}_{g}\bold{A}_{g} \) serve as cross-channel transformations. The function \( f(\cdot) \) denotes an embedded Gaussian function. \( C(\bold{X}^{k}) \) functions as a normalization factor. For a given \( i \) and \( k \), the expression \( \frac{1}{C(\bold{X}^{k})}\sum_{\forall j} f(\bold{X}^{k}_{i}, \bold{X}^{k}_{j}) \) denotes the softmax computation along the \( j \)-dimension in the \( k \)-th window. Ultimately, a residual operation is applied in this attention mechanism. The resulting output is as follows:
\begin{equation}
    \begin{aligned}
        \bold{Z}^{k}_{i} = \bold{W}_{z} \bold{Y}^{k}_{i} + \bold{X}^{k}_{i} + \Delta \bold{W}_{z} \bold{Y}^{k}_{i} + \bold{X}^{k}_{i}.
    \end{aligned}
\end{equation}
The matrices \( \bold{W}_z \) and \( \Delta \bold{W}_{z} = \bold{B}_{z}\bold{A}_{z} \) represent the weight matrices used for computing the position-wise embedding on \( \bold{Y}^{k}_{i} \), as described in Equation \ref{eq6}.

\subsection{Swin-Transformer Block with LoRA}
STB plays a vital role in the STF image compression model because it integrates window-based multi-head self-attention (W-MSA) and shifted window-based multi-head self-attention (SW-MSA), effectively capturing long-range redundancy for improved information embedding. Each (S)W-MSA requires an input: the image embedding \( \bold{I} \) \(\in \mathbb{R}^{HW \times C} \), initially reshaped to \( \bold{I} \) \(\in \mathbb{R}^{H \times W \times C} \). Then, the input matrix \( \bold{I} \) is partitioned into windows of size \( d \), followed by unfolding each window along the token dimension. This results in \( \bold{I} \) \( \in \mathbb{R}^{N \times d^2 \times C} \), where \( N = \frac{H \times W}{d^2} \). The self-attention mechanism for a specific head within a window is illustrated below:
\begin{equation}
    \begin{aligned}
        Attention(\bold{Q}, \bold{K}, \bold{V}) = Softmax(\bold{Q} \bold{K} / \sqrt{C} + \bold{B})\bold{V}.
    \end{aligned}
\end{equation}

In this formulation, \( \bold{Q} = \bold{I} (\bold{W}_{Q} + \Delta \bold{W}_{Q}) \), \( \bold{K} = \bold{I} (\bold{W}_{K} + \Delta \bold{W}_{K}) \), and \( \bold{V} = \bold{I} (\bold{W}_{V} + \Delta \bold{W}_{V}) \). Here, \( \bold{I} \in \mathbb{R}^{d^2 \times C} \) represents the window unfolding of the image embedding. \( \bold{W}_{Q} \), \( \bold{W}_{K} \), \( \bold{W}_{V} \in \mathbb{R}^{C \times C} \), and \( \bold{B} \in \mathbb{R}^{d^2 \times d^2} \) are pre-trained parameters that remain fixed, while \( \Delta \bold{W}_{Q} = \bold{B}_{Q}\bold{A}_{Q} \), \( \Delta \bold{W}_{K} = \bold{B}_{K}\bold{A}_{K} \), \( \Delta \bold{W}_{V} = \bold{B}_{V}\bold{A}_{V} \) are trainable parameters. Afterward, we conduct the window reverse and resize operation, resulting in (S)W-MSA outputs \( I' \in \mathbb{R}^{HW \times C} \). Another important component in the Transformer is called the multi-layer perceptron (MLP).We further integrate LoRAM into this MLP as \( \bold{O} = GELU((\bold{I}\bold{W}_{FC_{1}} + \Delta \bold{W}_{FC_{1}})) (\bold{W}_{FC_{2}} + \Delta \bold{W}_{FC_{2}})\  \). In this formulation, only \( \Delta \bold{W}_{FC_{1}} = \bold{B}_{FC_{1}}\bold{A}_{FC_{1}}  \), \( \Delta \bold{W}_{FC_{2}} = \bold{B}_{FC_{2}}\bold{A}_{FC_{2}} \) are trainable parameters.

\subsection{Layer Selection}

\begin{figure}[htbp]
\centerline{\includegraphics[scale=0.33, page=3]{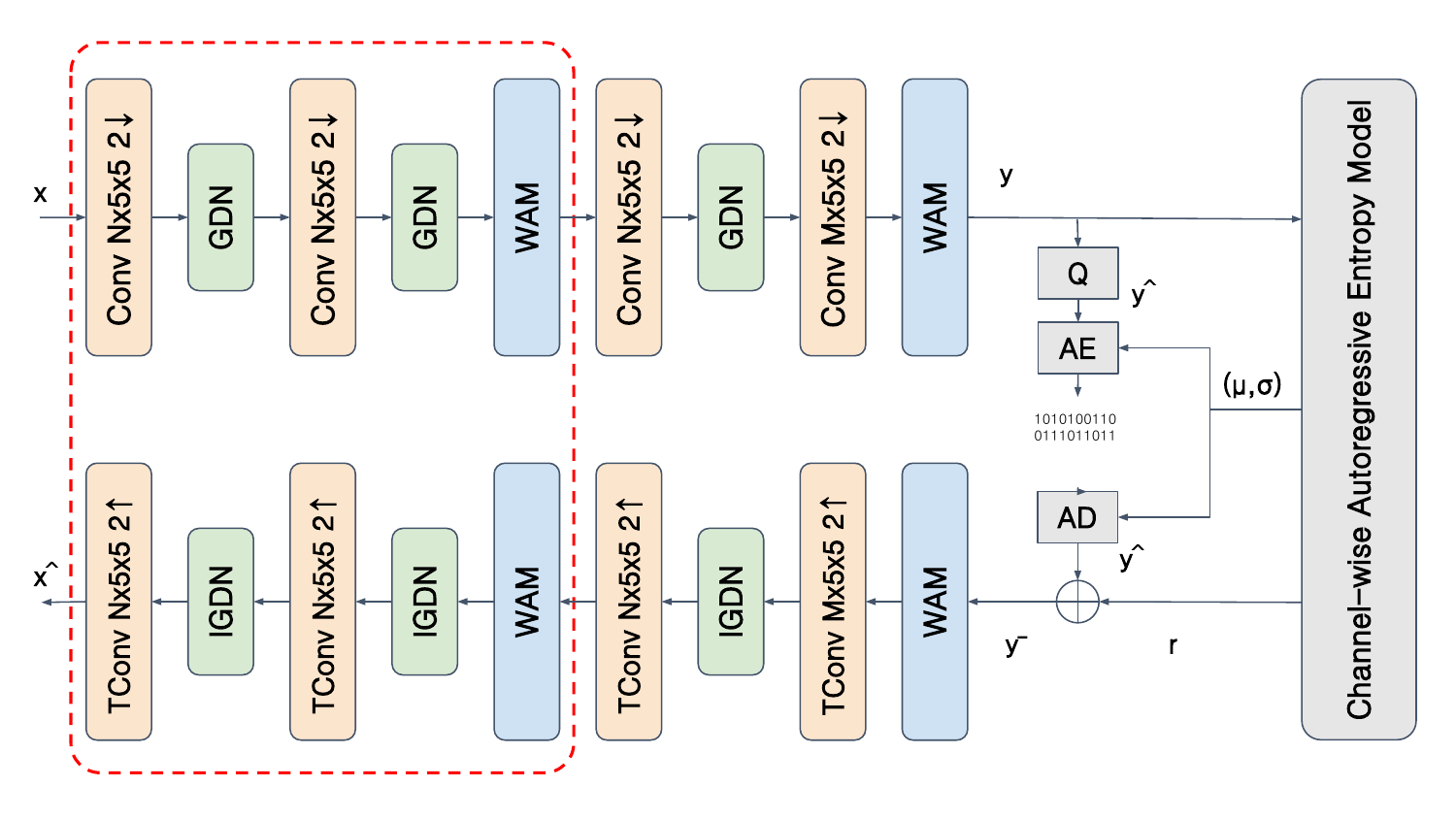}}
\caption{The layer selection of WACNN for integrating LoRAM into our network.}
    \label{fig:Our_WACNN}
\label{fig}
\end{figure}

\begin{figure}[htbp]
\centerline{\includegraphics[scale=0.33, page=4]{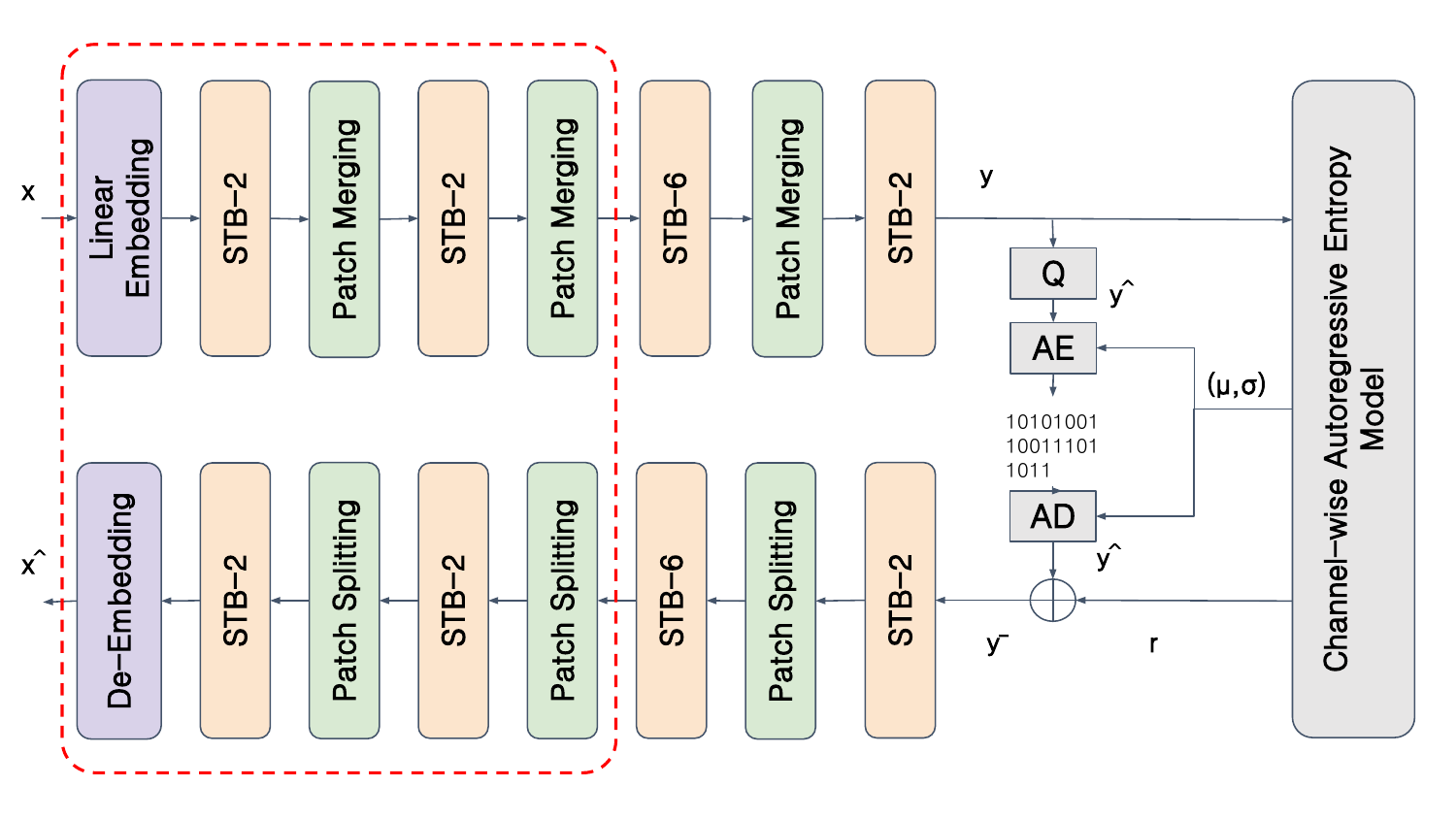}}
\caption{The layer selection of STF for integrating LoRAM into our network.}
    \label{fig:Our_STF}
\label{fig}
\end{figure}

In recent years, deep neural networks have demonstrated exceptional performance across various imaging tasks. Consequently, when assessing the effectiveness of image compression, it is essential to evaluate not only based on human perception but also the impact of compression on downstream tasks. In deep neural networks, shallower layers are typically responsible for capturing low-level image details. Therefore, we apply LoRAM to these shallower layers, allowing us to achieve similar performance with fewer additional model parameters. We integrate LoRAM into all layers within the regions enclosed by red dashed lines, as illustrated in Figs. \ref{fig:Our_WACNN} and \ref{fig:Our_STF}.

\subsection{Variable Rate-distortion Loss}
The objective of deep image compression is to minimize the rate-distortion function, representing the balance between the length of the bitstream and the distortion between the reconstructed image and the origin image. In our approach, we initially pre-train a model for a target rate corresponding to a specific Lagrange multiplier \( \lambda_0 \).

\( R_{\tau}( \cdot ) \) denotes the expected code length (bitrate) of the quantized latent feature, while \( D_{\tau}( \cdot , \cdot ) \) quantifies the distortion between the origin and reconstructed images. Upon acquiring well-trained parameters \( {\phi_0} \), \( {\theta_0} \), \( {\omega_0} \), \( {\psi_0} \) and \( {\xi_0} \) for the backbone network, we incorporate the LoRA Rate-Adaptive Module, \( {\Delta \phi_i} = \{\bold{B}_{\phi_i} \), \( \bold{A}_{\phi_i}\} \), \( {\Delta \omega_i} = \{\bold{B}_{\omega_i} \), \( \bold{A}_{\omega_i}\} \), \( {\Delta \theta_i} = \{\bold{B}_{\theta_i} \), \( \bold{A}_{\theta_i}\} \), \( {\Delta \psi_i} = \{\bold{B}_{\psi_i} \), \( \bold{A}_{\psi_i}\}\) and \( {\Delta \xi_i} = \{\bold{B}_{\xi_i} \), \( \bold{A}_{\xi_i}\}\), and fine-tune the Lagrange multiplier \( \lambda_i \):
\begin{equation}
    \begin{aligned}
        \mathcal{L}_i &= R_{\tau}(\hat{y}, \hat{z}) + \lambda_i \cdot D_{\tau}(x, \hat{x})\\
        &= \mathbb{E}_{x\sim p_x}[-log_2 p_{\hat{y}|\hat{z}}(\hat{y}|\hat{z}) -log_2 p_{\hat{z}}(\hat{z})] \\ &\quad + \lambda \cdot \mathbb{E}_{x\sim p_x}[d(x, \hat{x}],\\
        \text{where}\quad &y = \mathcal{G}_{a}(x;{\phi_0}, {\Delta \phi_i}), \quad z = \mathcal{H}_{a}(y;{\omega_0}, {\Delta \omega_i}), \\
                    &\hat{y} = Q(y),\quad
                    \hat{z} = Q(\mathcal{H}_a(y;{\omega_0},{\Delta \omega_i})),\\
                    &\hat{x} = \mathcal{G}_s(\hat{y};{\theta_0}, {\Delta \theta_i}), \quad
                    p_{\hat{y}|\hat{z}}(\hat{y}|\hat{z}) 
                        = \mathcal{N}(\mu, \sigma^2), \\
                    &\mu, \sigma 
                    = \mathcal{G}_{ep}(
                            \mathcal{H}_{s}(
                                \hat{z};
                                {\psi_0}, {\Delta \psi_i}
                            )
                            ; {\xi_0}, {\Delta \xi_i}
                        ).
    \end{aligned}
\end{equation}

Upon training the parameters \( {\Delta \phi_i}\), \( {\Delta \omega_i}\), \( {\Delta \theta_i}\), \( {\Delta \psi_i}\) and \( {\Delta \xi_i}\), we derive variable compression models customized for the i-th target rate.

\section{Experiments}

\subsection{Experimental Setup}
Based on the original author's implementation, we realize our proposed LoRAM into WACNN and STF. The training process consists of two stages. 

During the initial phase, we fix the Lagrange multiplier \(\lambda_0=0.025\) and exclusively optimize the parameters of the backbone network, including \( \mathcal{G}_{a}(\cdot; \phi) \), \( \mathcal{H}_a(\cdot; \omega) \), \( \mathcal{G}_s(\cdot; \theta) \), \( \mathcal{H}_s(\cdot; \psi) \), and \( \mathcal{G}_{ep}(\cdot, \xi) \). We randomly sample 300 k images from the OpenImages dataset \cite{b14} and crop them to a size of 256 × 256. The backbone model undergoes training for 1.8M iterations employing a batch size 16 and a learning rate \( 1 \times 10^{-4} \), utilizing the Adam optimizer \cite{b13}.

In the second stage, we train the LoRAM \(\Delta\phi_{i}\), \(\Delta\omega_{i}\), \(\Delta\theta_{i}\), \(\Delta\psi_{i}\), and \(\Delta\xi_{i}\) separately for each Lagrangian operator, by setting \(\lambda_i\) to \{0.0018,0.0035,0.0067,0.013,0.0483\}. We randomly choose 32 k images from the OpenImages dataset \cite{b14} and crop them to a size of 256 × 256. The separate models are trained for 40 k steps with a batch size 16 and a learning rate \( 5 \times 10^{-4} \), using the Adam optimizer \cite{b13}.

We assess the performance of our method on WACNN and STF \cite{b32} models by computing the average rate-distortion (in terms of PSNR) on the widely utilized Kodak image set \cite{b10} and the CLIC professional validation dataset \cite{b29}. To assess the impact on the classification task, we create a test set from ImageNet \cite{b9} by randomly selecting 10 images from each category.

In WACNN, we configure 320 latent channels and 192 hyper latent channels, while for STF, we designate a patch size of 2 x 2, a window size of 4 x 4, and \( C \) channels set to 48. In general, all models maintain consistent hyperparameters across varying \( \lambda \) values. For both WACNN and STF, we integrate LoRAM into all layers within the regions enclosed by red dashed lines, as illustrated in Figures \ref{fig:Our_WACNN} and \ref{fig:Our_STF}. All convolutional layers utilize our LoRAM for the Channel-wise Autoregressive Entropy Model.

\subsection{Efficiency Comparison}
To evaluate the effect of integrating LoRAM on the parameters and training efficiency of image compression network models, Table \ref{tbl:params_cmp} provides a comparison of storage and training efficiency. This comparison underscores the benefits of LoRAM in terms of parameter storage and training efficiency. It shows that LoRAM requires merely 0.20M additional parameters, which is notably less than the 99.85M required for standalone STF \cite{b32}, and only 0.156M additional parameters, significantly less than the 75.23M needed for standalone WACNN \cite{b32}. Notably, when adjusting for a new target rate, only saving LoRAM is sufficient, unlike STF or WACNN \cite{b32}, which demand storing an entirely new model. This disparity results in a substantial 99\% reduction in storage space.
\begin{table}[htbp]
\caption{The comparison of the changes in network parameter quantities.}
\begin{center}
\begin{tabular}{|c|c|}
\hline
Variant & Params (M) \\\hline
STF$_{w/o\ LoRA}$ & 99.85 \\\hline
STF$_{w\ LoRA}$ & 99.85 + 0.2 ( +0.2\% ) \\\hline
WACNN$_{w/o\ LoRA}$ & 75.23 \\\hline
WACNN$_{w\ LoRA}$ & 75.23 + 0.156 ( +0.2\% ) \\\hline
\end{tabular}

\label{tbl:params_cmp}
\end{center}
\end{table}
Our method yields satisfactory results regarding training efficiency using only 10.6\% of the original dataset. It reaches convergence in just 40k steps, representing a significant improvement compared to STF or WACNN \cite{b32}, which require 1.8M steps, as shown in Table \ref{tbl:training_cmp}.

\begin{table}[htbp]
\caption{The comparison of the training dataset sizes and the number of training steps.}
\begin{center}
\centering
\begin{tabular}{|c|c|c|}
\hline
Variant & Training Dataset Size (k) &  Training Steps (M) \\\hline
STF & $300$ & $1.8$ \\\hline
WACNN & $300$ & $1.8$ \\\hline
Ours & $\bold{32}$ & $\bold{0.04}$ \\\hline
\end{tabular}
\label{tbl:training_cmp}
\end{center}
\end{table}

\subsection{Rate-Distortion Performance}
\begin{figure}[htbp]
\centerline{
    \includegraphics[scale=0.27, page=1]{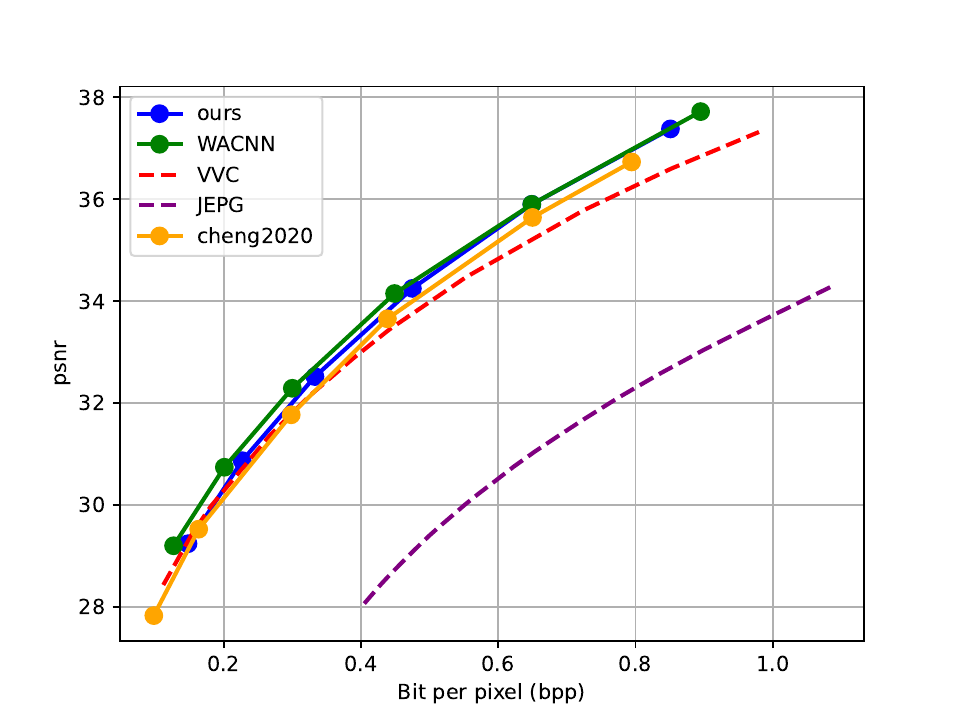}
    \includegraphics[scale=0.27, page=1]{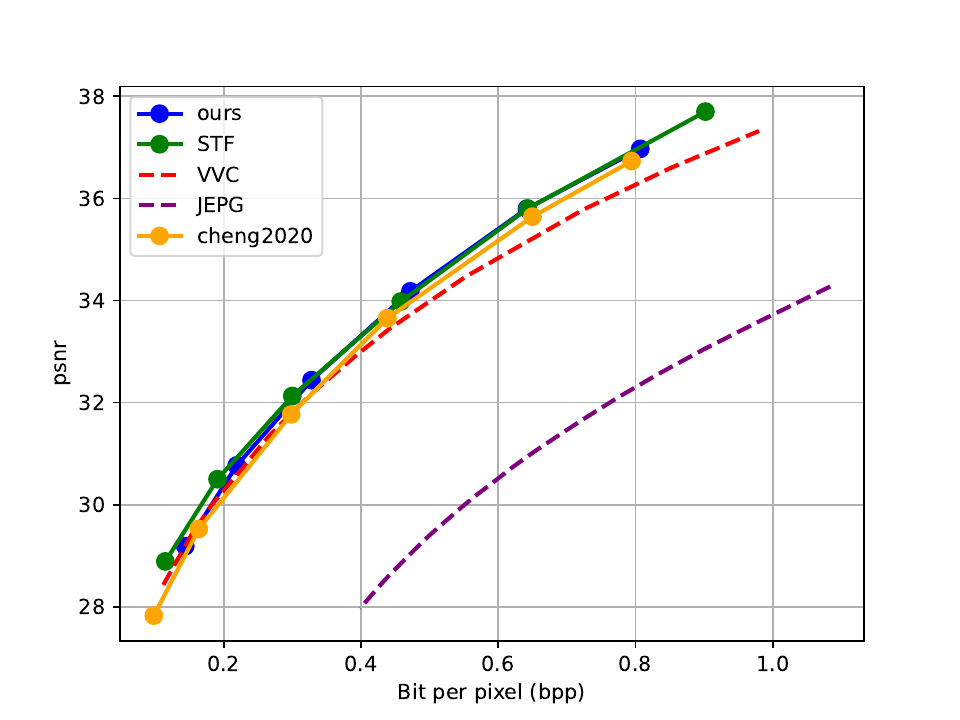}
    }
\caption{ The RD performance comparison on the Kodak dataset.}
    \label{fig:RD_comp}
\label{fig}
\end{figure}

\begin{figure}[htbp]
\centerline{
    \includegraphics[scale=0.27, page=1]{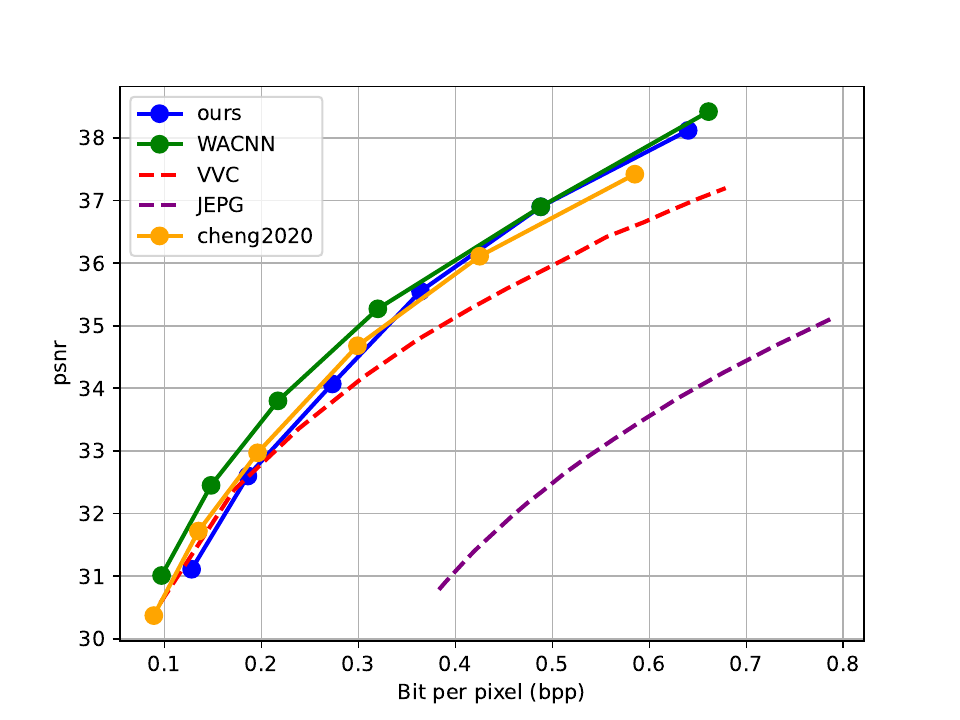}
    \includegraphics[scale=0.27, page=1]{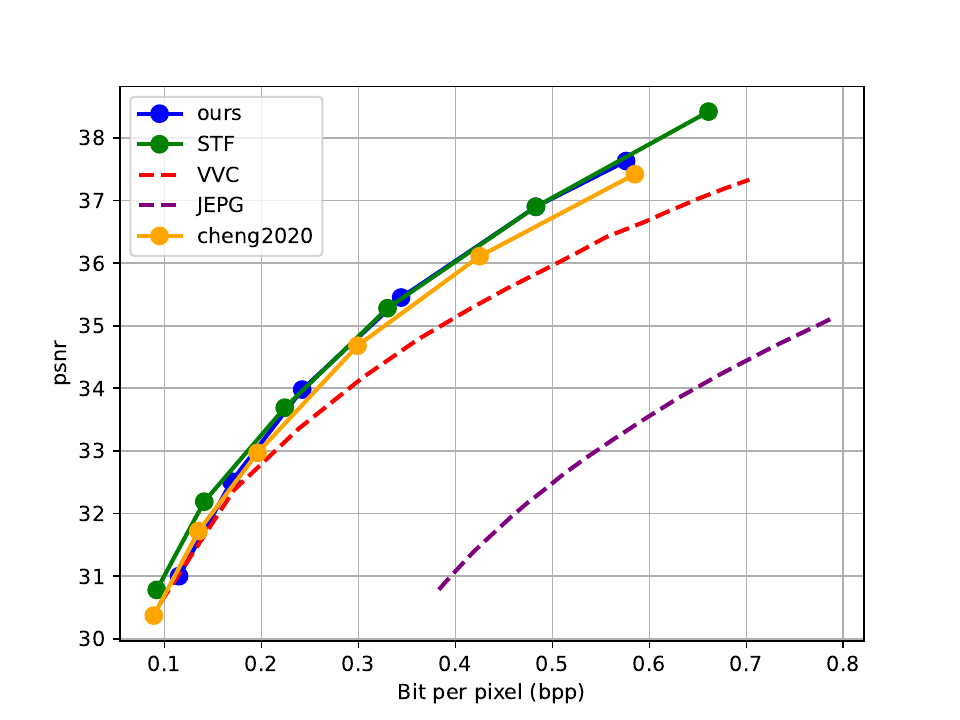}
    }
\caption{ The RD performance comparison on the CLIC dataset.}
    \label{fig:RD_CLIC_comp}
\label{fig}
\end{figure}

In this subsection, we assess the rate-distortion performance of this approach. Fig. \ref{fig:RD_comp} and Fig. \ref{fig:RD_CLIC_comp} illustrate the rate-distortion performance using the Kodak dataset \cite{b10} and the CLIC2020 dataset \cite{b29}, with PSNR as the evaluation metric. Our model is compared with the baseline models, WACNN \cite{b32}, STF \cite{b32}, the model proposed by Cheng et al. \cite{b6} and the classical image compression codec. Compared to the baseline models \cite{b32}, ours achieves comparable rate-distortion (R-D) performance with significantly fewer parameters, while avoiding performance degradation. In particular, our approach surpasses the widely adopted classical image codec JPEG \cite{b23} and demonstrates comparable performance to VTM \cite{b5} in PSNR. VTM is acknowledged as a leading intra-frame encoding method within the upcoming compression standard Versatile Video Coding (VVC) \cite{b5}. The performance of WACNN, when incorporating our method on the CLIC dataset, shows a slight degradation. This might be attributed to differences in image sizes between the CLIC dataset and the training set. Specifically, the convolution layers in WACNN are proficient at capturing short-range correlations, thus making them more susceptible to the influence of image size discrepancies.

\subsection{Classification Task Influence}

\begin{figure}[htbp]
\centerline{
    \includegraphics[scale=0.27, page=1]{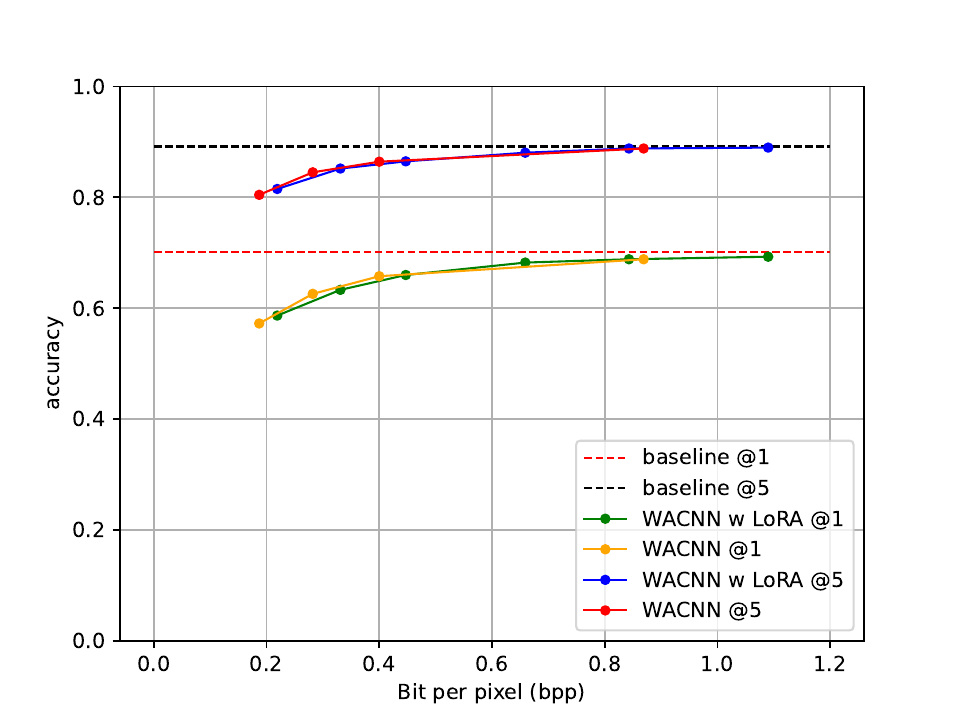}
    \includegraphics[scale=0.27, page=1]{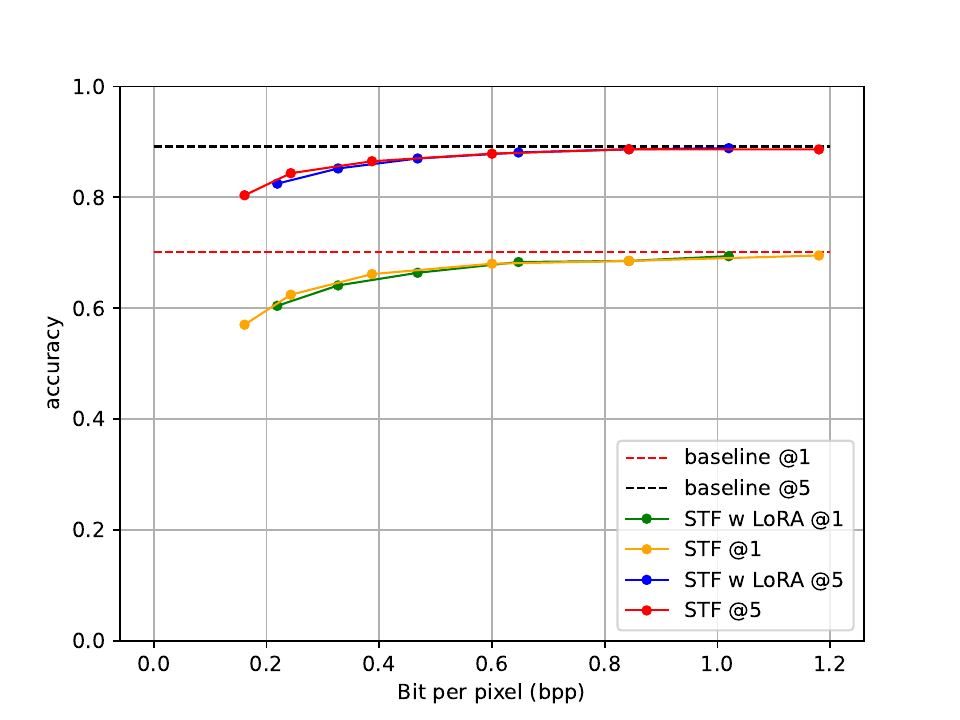}
    }
\caption{ Classification accuracy comparison on the ImageNet dataset. }
    \label{fig:Classification_comp}
\label{fig}
\end{figure}

Deep neural networks have demonstrated exceptional performance across various imaging tasks in recent years, leading to widespread adoption in numerous computer vision applications. It is crucial to evaluate the impact on the performance of these computer vision tasks while ensuring efficient data transfer and storage between machines running these data models. We assess the classification accuracy using a pre-trained ResNet18 model \cite{b11} to validate the classification performance. We randomly selected ten images from each category in the ImageNet1k dataset and evaluated the Top-1 and Top-5 accuracy of ResNet18 on these original images, using this as the baseline. Following this, we compressed and reconstructed these images using WACNN and STF at various compression rates. Subsequently, the Top-1 and Top-5 accuracies of ResNet18 were evaluated on the reconstructed images. Finally, we compared the performance of models fine-tuned with our method against those fully trained separately. Fig. \ref{fig:Classification_comp} presents the comparison results of different approaches and baseline performance. The model's performance trained using our method for different compression rates is very similar to that of the model fully trained separately. However, our method significantly reduces computational time and model storage requirements.

\section{Conclusion}
In conclusion, our research provided an overview of leveraging deep neural networks for image compression, explored how various designs enhance model performance, and examined the challenges and significance of the Variable-Rate Deep Image Compression task. We also explained the importance of different methods to address these challenges. We incorporated LoRA into the model to tackle these issues with the proposed LoRA Rate-Adaptive Module (LoRAM), enabling image compression at different rates using only 0.1\% additional model parameters.

With excellent computational and storage efficiency, we experimentally verified the effectiveness of using LoRA for Variable-Rate Deep Image Compression. Given the increasing reliance on data transfer between machines in many applications, we investigated whether our method would negatively impact downstream tasks. Our experiments demonstrated that our approach performs strongly in downstream tasks without degradation.

\end{document}